\documentclass[a4paper,12pt]{article}
\usepackage{amssymb,amsthm,latexsym}

\newcommand{\med}[1]{\langle #1 \rangle}
\newtheorem{theorem}{Theorem}

\title{The replica symmetric region in the Sherrington-Kirkpatrick mean field
spin glass model. The Almeida-Thouless line}

\author{
Francesco Guerra\footnote{\
e-mail: {\tt francesco.guerra@roma1.infn.it}} \\
{\small {\itshape Dipartimento di Fisica, Universit\`a di Roma ``La Sapienza''}}
\\
{\small {\itshape INFN, Sezione di Roma1, Piazzale A. Moro 2, 00185 Roma, 
Italy}}
} 
\date{\today}
\begin{document}

\maketitle

\begin{abstract}
\noindent
In previous work, we have developed  a  simple method to study  the behavior of the
Sher\-ring\-ton-Kirkpatrick mean field spin glass model for high temperatures, or equivalently
for high external fields. The basic idea was to
couple two different replicas with a quadratic term, trying to push out the two
replica overlap from its replica symmetric value. In the case of zero external
field, our results reproduced the well known validity of the annealed
approximation, up to the known critical value for the temperature. In the case
of nontrivial external field, our method could  prove the validity of 
the Sher\-ring\-ton-Kirkpatrick replica symmetric solution up to a line, which
fell short of the Almeida-Thouless line, associated to the onset of the
spontaneous replica symmetry breaking, in the Parisi Ansatz. Here, we make a strategic improvement of the method, by modifying the flow equations, with  respect to the parameters of the model. We exploit also previous results on the overlap fluctuations in the replica symmetric region. As a result, we give a simple proof that replica symmetry holds up to the critical Almeida-Thouless line, as expected on physical grounds. Our results are compared with the characterization of the replica symmetry breaking line previously given by Talagrand. We outline also a possible extension of our methods to the broken replica symmetry region.
\end{abstract}

\section{Introduction}

It is very well known, on physical grounds, that for the mean field spin glass model, introduced by
Sherrington and Kirkpatrick in \cite{SK}, the replica symmetric
solution holds in a region of the parameters characterized by high temperature and/or high external magnetic field, as shown for example in \cite{MPV}, and references quoted
there. Recently, Michel Talagrand \cite{T} has been able to give a complete characterization of this region, through a method which exploits the broken replica symmetry bounds, established in 
\cite{Grepli}, extended to the case of two coupled replicas. Talagrand method is remarkable, not only because it gives a full characterization of the replica symmetric region, but also because it can be easily extended, with some technical effort, to the broken replica symmetry region. In this way, it has been possible to establish the rigorous validity of the Parisi Ansatz for the free energy density in the infinite volume limit \cite{Topus}.

The motivation of this paper can be easily explained. It is surely true that some features of the broken replica symmetry phenomenon can be found also in the region of the parameters where replica symmetry still holds. This is a general aspect of statistical physics. Let us consider for example a system in its liquid phase near to the onset of crystallization. While the liquid phase is translation and rotation  invariant, one can see features of the crystal phase, where translation and rotation invariance are broken, through a study of the correlations, near the transition line. In the approach of Talagrand, the broken replica symmetry features, for coupled replicas, are exploited in an essential way in order to characterize the boundaries of the replica symmetric region. However, it should be possible, in principle, to characterize the replica symmetric region in terms of its own physical content, without resorting to features of the broken replica symmetry phenomenon. We will show that this is possible indeed.

 Our method is very simple. It can be considered a refinement of the interpolation method introduced in \cite{GTquadratic}. We will exploit also in a crucial way the established fact that overlap fluctuations are Gaussian in the replica symmetric region, as shown for example in \cite{GTcentral} and \cite{T}.
 
 The paper is organized as follows. In section 2 we recall the main features of our improved interpolation method. We consider an abstract general setting, and derive upper and lower bounds for the generalized free energy. These bounds are the essential ingredients to be exploited for the rigorous study of mean field spin glass models. In section 3 we introduce the mean field spin glass model in the replica symmetric region, and state the main results. The quadratic replica coupling for two replicas is introduced in section 4, and the main features of the corresponding replica symmetric region are investigated. Section 5 is devoted to the calculation of overlap fluctuations for systems containing two coupled replicas. These calculations follow the pattern outlined in \cite{Gsum} in the case of a single replica, and are a very simple extension of the treatment given in \cite{GTcentral}. The main results of this paper are proven in section 6, where it is shown in particular that the replica symmetric region extends up to the Almeida-Thouless line, as expected. We make also the comparison with the replica symmetric region given by Talagrand. In particular, we are able to isolate, in full generality, the parameter ruling the validity of a generic Ansatz for the free energy density. Section 7 is dedicated to some conclusion and outlook for future developments. 
 
 \section{Upper and lower bounds from interpolation methods}
 
For the sake of future reference, we give here a complete and self-contained treatment. Let us  introduce the following general setting. Let $U_i$ be a family
of centered Gaussian random variables, $i=1,\dots,K$, with covariance 
matrix given by $E(U_i U_j) \equiv S_{ij}$. We treat the set of indices $i$ now as 
configuration space for some statistical mechanics system, with partition 
function $Z$ and quenched 
free energy given by
\begin{equation}\label{frU}
E\log\sum_i w_i \exp (\sqrt{t} U_i)\equiv E\log Z,
\end{equation}
where $w_i\ge0$ are generic weights, and $t$ is a parameter ruling the 
strength of the interaction.

The following well known derivation 
formula plays a leading role in the interpolation method
\begin{eqnarray}
\nonumber
&& {d\over dt}E\log\sum_i w_i \exp (\sqrt{t} U_i)=\\
\nonumber
&& {1\over2}E(Z^{-1}\sum_i w_i \exp (\sqrt{t} U_i) S_{ii}\\
\label{derivation}
&& -{1\over2}
E(Z^{-2}\sum_i \sum_j w_i w_j \exp (\sqrt{t} U_i) \exp (\sqrt{t} U_j) 
S_{ij}).
\end{eqnarray}
The proof is straightforward. Firstly we perform directly the $t$ 
derivative. Then, we notice that the random variables appear in expressions 
of the form $E(U_i F)$, were $F$ are functions of the $U$'s. These can be 
easily handled through the following integration by parts formula for generic 
Gaussian random variables, analogous to the Wick theorem in 
quantum field theory,
\begin{equation}
\label{wick}
E(U_i F)=\sum_j S_{ij} E({\partial\over\partial{U_j}}F).
\end{equation}
Therefore, we see that always two derivatives are involved. The two terms 
in (\ref{derivation}) come from the action of the $U_j$ derivatives, the 
first acting on the Boltzmann factor, and giving rise to a Kronecker 
$\delta_{ij}$, the second acting on $Z^{-1}$, and giving rise to the minus 
sign and the duplication of variables. 

The derivation formula can be 
expressed in a more compact form by introducing replicas and suitable 
averages. In fact, let us introduce the state $\omega$ acting on 
functions $F$ of $i$ as follows
\begin{equation}\label{state}
\omega(F(i))=Z^{-1}\sum_i w_i \exp (\sqrt{t} U_i) F(i),
\end{equation}
together with the associated product state $\Omega$ acting on replicated 
configuration spaces $i_1,i_2,\dots,i_s$. By performing also a global $E$
average, finally we define the averages
\begin{equation}\label{bracket}
\med{F}_t\equiv E\Omega(F),
\end{equation}
where the subscript is introduced in order to recall the $t$-dependence of 
these averages.

Then, the equation (\ref{derivation}) can be written in a more compact form
\begin{equation}\label{derivation'}
{d\over dt}E\log\sum_i w_i \exp (\sqrt{t} U_i)=
{1\over2}\med{S_{i_1 i_1}}_t-{1\over2}
\med{
S_{i_1 i_2}}_t.
\end{equation}

By following the same procedure, we can also easily establish the following general derivation formula for $\med{}$ averages
\begin{equation}\label{derivation''}
{d\over dt}\med{F(i_1)}_t=
{1\over2}\med{F_{i_1} (S_{i_1 i_1}-S_{i_2 i_2}-2S_{i_1 i_2}+2S_{i_1 i_3})}_t.
\end{equation}
Here, up to three replicas are involved. In fact, the first $t$ derivative will let a second replica appear, while the subsequent $\partial/\partial A_j$ derivatives will introduce also a third replica.

Let us now go to the interpolation and comparison arguments.
Let $U_i$ and $\hat U_i$, for $i=1,\dots,K$, be independent families of 
centered Gaussian random variables, with covariance matrices given by  $E(U_i U_j) \equiv S_{ij}$ and $E(\hat U_i \hat U_j) \equiv \hat S_{ij}$. For $0 \le t \le 1$, introduce the interpolating function 
\begin{equation}\label{alfa}
\alpha(t)\equiv E\log\sum_i w_i \exp (\sqrt{t} U_i) \exp (\sqrt{1-t} \hat U_i),
\end{equation}
so that we have the boundary terms
 \begin{equation}\label{alfa0}
\alpha(0)\equiv E\log\sum_i w_i  \exp ( \hat U_i)
\end{equation}
\begin{equation}\label{alfa1}
\alpha(1)\equiv E\log\sum_i w_i  \exp ( U_i).
\end{equation}  

By applying the $t$ derivative as before, and taking into account that the $1-t$ term will bring an extra minus sign, we arrive to the basic interpolation formula
\begin{equation}\label{dalfa}
{d\over dt}\alpha(t)=
{1\over2}\med{S_{i_1 i_1}-\hat S_{i_1 i_1}}_t-{1\over2}
\med{S_{i_1 i_2}-\hat S_{i_1 i_2}}_t.
\end{equation}
In the simplest case, the comparison argument runs as follows. Assume that the
covariances satisfy the 
inequalities for generic configurations
\begin{equation}\label{dominance}
E(U_i U_j) \equiv S_{ij} \ge E(\hat U_i \hat U_j) \equiv \hat S_{ij},
\end{equation}
and the equalities along the diagonal
\begin{equation}\label{diagonal}
E(U_i U_i) \equiv S_{ii} = E(\hat U_i \hat U_i) \equiv \hat S_{ii},
\end{equation}
then we have
\begin{equation}\label{dalfaless}
{d\over dt}\alpha(t)\le 0,
\end{equation} 
with the obvious consequences
\begin{equation}\label{upperbound}
\alpha(1)\le\alpha(0),\ \ \ \alpha(t)\le\alpha(0).
\end{equation}
In the applications, $\alpha(1)$ is the quantity of interest, while $\alpha(0)$ is connected to some comparison system, and can be easily calculated. Therefore, we call (\ref{upperbound}) an upper bound formula.

Considerations of this kind are present in the mathematical 
literature of some years ago. Two typical references are \cite{Joag} 
and \cite{Kahane}.

If the stated conditions for the covariances (\ref{dominance},\ref{diagonal}) hold, then we also have
\begin{equation}\label{dalfa'}
{d\over dt}(\alpha(0)-\alpha(t))=
{1\over2}\med{\Delta_{12}}_t,
\end{equation} 
where
\begin{equation}\label{Delta}
\Delta_{12}\equiv S_{i_1 i_2}-\hat S_{i_1 i_2}\ge 0.
\end{equation} 
By integration, we have also
\begin{equation}\label{sumrule}
\alpha(0)=\alpha(1)+
{1\over2}\int_0^1\med{\Delta_{12}}_t dt.
\end{equation} 
We call ``sum rule'' an expression of this kind, by following the terminology introduced in \cite{Gsum}.

The derivation formula and the comparison argument are not restricted to the 
Gaussian case. Generalizations in many directions are possible. For the 
diluted spin glass models and optimization problems we refer for example 
to \cite{FL},  and to 
\cite{LDS}, and references quoted there.

Let us show how to derive lower bounds, in the opposite direction. We follow the general strategy expounded in \cite{GTquadratic}, with some relevant improvement.

Introduce a system made by two replicas with some appropriate coupling between them. Now the interpolating quenched free energy is defined as
\begin{eqnarray}
\nonumber
&&\tilde\alpha(\lambda,t)\equiv\\
\nonumber
&&{1\over2}E\log\sum_{ij} w_i  w_j \exp (\sqrt{t} (U_i+U_j)+\sqrt{1-t} (\hat U_i+\hat U_j))\exp(\lambda(S_{ij}-\hat S_{ij})),\\
&&\label{alfatilde}
\end{eqnarray}
depending on two parameters, the interpolating parameter $0\le t\le1$, and the coupling parameter $\lambda\ge0$. We have also introduced the factor $1/2$ in order to keep the same overall normalization. In fact, we have $\tilde\alpha(0,t)=\alpha(t)$, if $\lambda$ is put to zero.

It is simple to calculate the $t$ derivative by a direct application of the general formula (\ref{dalfa}). Of course, now we have to take into account that the original system is made of two (coupled) replicas, let us say $(1,2)$. The replication process, therefore, will introduce a replicated system made by the replicas $(3,4)$, coupled in the same way as $(1,2)$. Now the diagonal term gives a nontrivial contribution, and we find
\begin{equation}\label{dtalfatilde}
\partial_t \tilde\alpha(\lambda,t)=
{1\over2}\med{\Delta_{12}}_{\lambda, t}-\med{\Delta_{13}}_{\lambda, t},
\end{equation}
where as before
\begin{equation}\label{Delta'}
\Delta_{12}\equiv S_{i_1 i_2}-\hat S_{i_1 i_2}\ge 0,\ \ \  \Delta_{13}\equiv S_{i_1 i_3}-\hat S_{i_1 i_3}\ge 0.
\end{equation}
The terms $\Delta_{14}, \Delta_{23}, \Delta_{24}$ do not appear explicitly, because we have exploited the $(1-2)$ and $(3-4)$ symmetries. Of course, because of the 
$\lambda$ coupling,  there is no symmetry of the type $(1-3)$, and analogous.
However, at $\lambda=0$, we have
\begin{equation}\label{lambda0}
 \med{\Delta_{12}}_{0, t} \equiv \med{\Delta_{12}}_t\equiv \med{\Delta_{13}}_{0, t}.
\end{equation}
 On the other hand, the $\lambda$ derivative is given simply by
 \begin{equation}\label{dlambdaalfatilde}
\partial_\lambda \tilde\alpha(\lambda,t)=
{1\over2}\med{\Delta_{12}}_{\lambda, t}.
\end{equation}

Introduce now a velocity field $v(\lambda,t)$, and trajectories $t\to\lambda(t)$, such that $\lambda(0)=\lambda_0$, and $\dot \lambda(t)=v(\lambda(t),t)$. Then we have for the total $t$ derivative
\begin{equation}\label{total}
\frac{d}{dt}  \tilde\alpha(\lambda(t),t)=
{1\over2}(1+v(\lambda(t),t))\med{\Delta_{12}}_{\lambda, t}-\med{\Delta_{13}}_{\lambda, t}.
\end{equation} 

In our previous paper \cite{GTquadratic}, we made the very simple choice 
$v(\lambda,t)=-1$, so that $\lambda(t)=\lambda_0-t$. In this case
\begin{equation}\label{total'}
\frac{d}{dt}  \tilde\alpha(\lambda(t),t)= -\med{\Delta_{13}}_{\lambda, t}\le 0.
\end{equation}
Therefore
\begin{equation}\label{tildeless}
\tilde\alpha(\lambda(t),t)\le \tilde\alpha(\lambda_0,0).
\end{equation}

A more efficient way is to go along the level lines of $\tilde\alpha$, with the natural choice
\begin{equation}\label{vbuona}
v(\lambda,t)=-1+2\frac{\med{\Delta_{13}}_{\lambda, t}}{\med{\Delta_{12}}_{\lambda, t}},
\end{equation}
so that
\begin{equation}\label{tildeequal}
\frac{d}{dt}  \tilde\alpha(\lambda(t),t)=0,\ \ \ \tilde\alpha(\lambda(t),t)\equiv \tilde\alpha(\lambda_0,0).
\end{equation}

This choice will play a central role in our treatment. Notice that $v(0,t)\equiv 1$. Let us now derive the lower bound. From the definition (\ref{alfatilde}) of $\tilde\alpha$, by applying Jensen inequality, we easily get
\begin{eqnarray}
\nonumber
\tilde\alpha(\lambda(t),t)
&\equiv&\alpha(t)+\frac{1}{2}E\log \Omega(\exp(\lambda(t) \Delta_{12}))\\
\label{jensen}
&\ge&\alpha(t)+\frac{\lambda(t)}{2}\med{\Delta_{12}}_t.
\end{eqnarray}
By recalling (\ref{dalfa'}) and (\ref{tildeless}) or (\ref{tildeequal}), we obtain
\begin{equation}\label{ldt}
\lambda(t){d\over dt}(\alpha(0)-\alpha(t))\le
\tilde\alpha(\lambda_0,0)-\alpha(t).
\end{equation}  
Provided the condition $\lambda(t^\prime)>0$ holds for $0\le t^\prime\le t$, we can easily integrate (\ref{ldt}), and derive the basic lower bound for $\alpha(t)$
\begin{equation}\label{lowerbound}
0\le\alpha(0)-\alpha(t))\le
(\tilde\alpha(\lambda_0,0)-\alpha(0))(\exp(\int_0^t \frac{dt^\prime}{\lambda(t^\prime)})-1).
\end{equation}
In the following we will make extensive use of this lower bound.

\section{The mean field spin glass model in the replica symmetric region}

The generic configuration of the mean field spin 
glass model is defined 
through Ising spin variables
$\sigma_{i}=\pm 1$,  attached to each site $i=1,2,\dots,N$.

The Hamiltonian of the model has a random character, and can be defined as follows.
First of all let us introduce 
a family of centered Gaussian random variables ${\cal K}(\sigma)$, indexed by the  
configurations $\sigma$, and characterized by the covariances
\begin{equation}\label{cov}
E\bigl({\cal K}(\sigma){\cal K}(\sigma^\prime)\bigr)=q^2(\sigma,\sigma^\prime),
\end{equation}
where $q(\sigma,\sigma^\prime)$ are the overlaps between two generic 
configurations, defined by
\begin{equation}\label{overlap}
q(\sigma,\sigma^\prime)=
N^{-1}\sum_{i}\sigma_i\sigma^{\prime}_i,
\end{equation}
with the obvious bounds
$-1\le q(\sigma,\sigma^\prime)\le 1$, and the normalization 
$q(\sigma,\sigma) = 1$.

At a generic inverse temperature $\beta$, for an external magnetic field of strength $h$,  
the (random) partition function is defined in the form  
\begin{equation}\label{Zsg}     
Z_N(\beta,h;{\cal K})=
\sum_{\sigma_1\dots\sigma_N}\exp(\beta \sqrt{N\over2}{\cal K}(\sigma))
\exp(\beta h \sum_{i}\sigma_i),
\end{equation}    
which will be the starting point of our treatment.

By now standard methods \cite{GTthermo} assure the existence of the infinite volume limit in the form
\begin{equation}\label{lim}     
\lim_{N\to\infty} N^{-1} E\log Z_{N}(\beta,h;{\cal K})\equiv
\sup_{N} N^{-1} E\log Z_{N}(\beta,h;{\cal K})\equiv \alpha(\beta,h). 
\end{equation}

Now we introduce the Sherrington-Kirkpatrick replica symmetric solution, by following the methods explained in \cite{Gsum}.

For a generic trial parameter $\bar q$, and an interpolating $0\le t\le 1$, define now
\begin{eqnarray}\nonumber     
\alpha_{N}(t)&\equiv&
N^{-1}E\log\sum_{\sigma_1\dots\sigma_N}\exp(\beta \sqrt{t}\sqrt{N\over2}{\cal K}(\sigma))
\exp(\beta \sum_{i}(h+\sqrt{1-t}\sqrt{\bar q}J_{i})\sigma_{i})\\
\label{alfat}
&+&\frac{1}{2}\beta^{2}(\bar q -1)^{2}(1-t),
\end{eqnarray}
where we have introduced independent unit Gaussian fields $J_{i}$, acting as magnetic fields on each site. Now $E$ denotes average over the ${\cal K}$'s and the $J_{i}$'s.

For $t=1$ we have
\begin{equation}\label{alfa1'}     
\alpha_{N}(1)\equiv N^{-1} E\log Z_{N}(\beta,h;{\cal K}),
\end{equation}
while for $t=0$, the \textit{Boltzmannfaktor} factorizes, the sum over $\sigma$'s can be performed explicitly, and we obtain  the replica symmetric trial function
\begin{equation}\label{alfa0'}     
\alpha_{N}(0)\equiv \log2 + \int \log\cosh\beta(h+\sqrt{\bar q}z) d\mu(z) + \frac{1}{2}\beta^{2}(\bar q -1)^{2}.
\end{equation}
Here $d\mu(z)$ is the unit Gaussian measure. Of course, $\alpha_{N}(0)$ does not depend on $N$, and will be written as $\alpha(0)$ in the following.

By the same procedure as in Section 2, we can easily calculate the $t$ derivative in the form
\begin{equation}\label{dalfa''}
\frac{d}{dt}\alpha_{N}(t)=-\frac{\beta^{2}}{4}\med{(q_{12}-\bar q)^{2}}_{t}\le 0.
\end{equation}
Clearly we have the uniform bounds
\begin{equation}\label{bounds}
\alpha_{N}(t)\le\alpha(0),\ \ \ \alpha_{N}(1)\le\alpha(0). 
\end{equation}
In view of these bounds, since $\bar q$ is a free parameter, it is convenient to take into (\ref{alfa0'}) the value that minimizes $\alpha(0)$. This value $\bar q(\beta,h)$ is uniquely defined, as shown in \cite{Gsum}, and satisfies the implicit equation
\begin{equation}\label{qbar}
\bar q(\beta,h)=\int\tanh^{2}(\beta(h+\sqrt{\bar q(\beta,h)}z)\ d\mu(z).
\end{equation}
It turns out that $\bar q(\beta,h)>0$, unless $h=0$ and $\beta\le1$ (the annealed region), where $\bar q(\beta,0)=0$. From now on, we assume that this minimizing value  $\bar q(\beta,h)$ has been chosen in all our formulae for $\bar q$. In this case $\alpha(0)$ in (\ref{alfa0'}) gives the replica symmetric Sherrington-Kirkpatrick value $\alpha_{SK}(\beta,h)\equiv\alpha(0)$, and we have
\begin{equation}\label{dalfaSK}
\frac{d}{dt}(\alpha_{SK}(\beta,h)-\alpha_{N}(t))=\frac{\beta^{2}}{4}\med{(q_{12}-\bar q)^{2}}_{t}\ge 0.
\end{equation}
Now the bounds in (\ref{bounds}) are given in the optimal form
\begin{equation}\label{bounds'}
\alpha_{N}(t)\le\alpha_{SK}(\beta,h),\ \ \ \alpha_{N}(1)\le\alpha_{SK}(\beta,h). 
\end{equation}

Let us now introduce the parameter $y_{0}\equiv y_{0}(\beta,h)$ defined by
\begin{equation}\label{y0}
y_{0}(\beta,h)\equiv\int\cosh^{-4}(\beta(h+\sqrt{\bar q(\beta,h)}z)) d\mu(z).
\end{equation}
On the $(\beta,h)$ plane, $\beta\ge0$, $h\ge0$, define the SK region, the AT line and the P region as the regions where $\beta^{2}y_{0}(\beta,h)\le1$, $\beta^{2}y_{0}(\beta,h)=1$ and $\beta^{2}y_{0}(\beta,h)\ge1$, respectively.

The main result of this paper is contained in the following theorem.
\begin{theorem}\label{main} 
In the replica symmetric region SK, where $\beta^{2}y_{0}(\beta,h)\le1$, the Sherrington-Kirkpatrick value gives the right free energy in the infinite volume limit, \textit{i.e.} 
\begin{eqnarray}\nonumber     
&&\lim_{N\to\infty} N^{-1} E\log Z_{N}(\beta,h;{\cal K})\equiv \alpha(\beta,h)\equiv\alpha_{SK}(\beta,h)\equiv\\
\label{limSK}
&&\inf_{\bar q}
(\log2 + \int \log\cosh(\beta(h+\sqrt{\bar q}z)) d\mu(z) + \frac{1}{2}\beta^{2}(\bar q -1)^{2}).
\end{eqnarray}
\end{theorem}

Let us now consider the interpolation function $\alpha_{N}(t)$ as defined in (\ref{alfat}). The methods explained in \cite{GTquadratic} show that there exists $\bar t(\beta,h)$, $0<\bar t(\beta,h)\le1$, such that for any $0\le t \le \bar t(\beta,h)$, we have in the limit
\begin{equation}\label{lim'}
\lim_{N\to\infty} \alpha_{N}(t)=\alpha_{SK}(\beta,h).
\end{equation}
We assume that $\bar t(\beta,h)$ is the largest value with this property, and call the interval $0\le t \le \bar t(\beta,h)$ the replica symmetric interval for $t$. Notice that this interval is always nontrivial, since $\bar t(\beta,h)$ is always larger than zero.

The following Theorem gives a full characterization of $\bar t(\beta,h)$.
\begin{theorem}\label{maint} 
In the replica symmetric region SK, where $\beta^{2}y_{0}(\beta,h)\le1$, $\bar t(\beta,h)$ takes its maximum value $\bar t(\beta,h)=1$. In the broken replica region, where $\beta^{2}y_{0}(\beta,h)>1$, the $\bar t(\beta,h)$ takes the value
\begin{equation}\label{tbar}
\bar t(\beta,h)=\beta^{-2}y_{0}(\beta,h)^{-1}.
\end{equation}
\end{theorem}

Clearly, Theorem \ref{main} follows from the first part of Theorem \ref{maint}, because we can take $t=1$ in (\ref{lim'}) and recall (\ref{alfa1'}).

On the other hand, the first part of Theorem \ref{maint} follows from the second part, by taking the limit $\beta^{2}y_{0}\to1$. The argument by Toninelli \cite{TAT}, based on the broken replica symmetry bounds of \cite{Grepli}, shows that, in the broken replica symmetry region, the parameter $\bar t(\beta,h)$ can not be greater than $\beta^{-2}y_{0}(\beta,h)^{-1}$. Therefore, in order to prove Theorem \ref{maint}, it is enough to show that $\bar t(\beta,h)$ can not be less than $\beta^{-2}y_{0}(\beta,h)^{-1}$. This will be our objective in the following.     

\section{Quadratic coupling of replicas}

Now we couple two replicated systems, with spin variables $\sigma^{(1)}$, $\sigma^{(2)}$, in the same spirit as in \cite{GTquadratic} and in Section 2. In analogy with (\ref{alfatilde}) the auxiliary function $\tilde \alpha_{N}(\lambda,t)$ is now defined, for $\lambda\ge0$ and $0\le t\le1$, as
\begin{eqnarray}
\nonumber
&&\tilde\alpha_{N}(\lambda,t)\equiv\frac{1}{2N}E\log\sum_{\sigma^{(1)}\sigma^{(2)}} \exp(B_{1})\exp(B_{2})\exp(\frac{1}{2}\lambda \beta^{2}N(q_{12}-\bar q(\beta,h))^{2})\\
\label{alfatilde'}
&&+\frac{1}{2}\beta^{2}(\bar q -1)^{2}(1-t),
\end{eqnarray}
where $B_{1}$ and $B_{2}$ are the two replicas of the quantities appearing in the \textit{Boltzmannfaktor} of $\alpha_{N}(t)$ in (\ref{alfat}).

The function $\tilde\alpha_{N}(\lambda,t)$ is convex increasing in $\lambda$ for each value of $t$, and we have
\begin{equation}\label{raccordo}
\tilde\alpha_{N}(0,t) \equiv \alpha_{N}(t),\ \ \ \tilde\alpha_{N}(\lambda,t)\ge\alpha_{N}(t).
\end{equation} 

The $t$ and $\lambda$ partial derivatives are easily calculated in analogy with (\ref{dtalfatilde}) and (\ref{dlambdaalfatilde})
\begin{equation}\label{dtalfatilde'}
\partial_t \tilde\alpha_{N}(\lambda,t)=
\frac{\beta^{2}}{4}\med{(q_{12}-\bar q)^{2}}_{\lambda, t}-\frac{\beta^{2}}{2}\med{(q_{13}-\bar q)^{2}}_{\lambda, t},
\end{equation}
\begin{equation}\label{dlambdaalfatilde'}
\partial_\lambda \tilde\alpha_{N}(\lambda,t)=
\frac{\beta^{2}}{4}\med{(q_{12}-\bar q)^{2}}_{\lambda, t}.
\end{equation}  

Let us go to the infinite volume limit $N\to\infty$, by exploiting for example the methods outlined in \cite{GTnonising}, and call $\tilde\alpha(\lambda,t)$ the resulting limit of $\tilde\alpha_{N}(\lambda,t)$. For any $0\le t\le \bar t(\beta,h)$, let $\bar \lambda(t)$ be the highest value of $\lambda$ for which $\tilde\alpha(\lambda,t)=\alpha_{SK}(\beta,h)$. We call the region $0\le t\le \bar t(\beta,h)$, $0\le\lambda\le\bar \lambda(t)$ the replica symmetric SK region in the $(\lambda,t)$ plane. 

In \cite{GTquadratic}, the value of $\bar\lambda(0)$ was calculated explicitly through a variational principle. In particular, in the case $h=0$, $\bar q=0$, it turns out that $\bar\lambda(0)=\beta^{-2}$. Moreover, it was shown that the triangular, or trapezoidal, region $0\le t\le\min(1,\bar\lambda(0))$, $0\le\lambda\le\bar\lambda(0)-t$, is contained into the replica symmetric region. Another interesting property of the replica symmetric region is the following. Assume that $(\lambda_{1},t_{1})$ is in the SK region, then all points $t_{1}\le t\le \min(1,t_{1}+\lambda_{1})$, $0\le \lambda\le\lambda_{1}-t+t_{1}$, are in the SK region. All these properties follow, according to \cite{GTquadratic}, from the bound (\ref{lowerbound}), adapted to the $\alpha_{N}(t)$ defined in (\ref{alfat}), with the simple choice $v(\lambda,t)\equiv-1$, by taking the infinite volume limit. 

A very important consequence is the following
\begin{theorem}\label{focus}
In the case where $\bar t(\beta,h)<1$, the following bound necessarily holds for $\bar\lambda(t)$
\begin{equation}\label{focus'}
\bar\lambda(t)\le \bar t(\beta,h)-t,
\end{equation}
for any $0\le t\le \bar t(\beta,h)$.
\end{theorem}

Therefore, if $\bar t(\beta,h)<1$, the replica symmetric region must shrink to the single point $\bar\lambda(\bar t(\beta,h))\equiv0$ at the end point $\bar t(\beta,h)$. The proof is straightforward. Should (\ref{focus'}) fail, there would be a point $t_{1}$ where  $\bar\lambda(t_{1})> \bar t(\beta,h)-t_{1}$. But in this case, by the general properties pointed before, it would be possible to enlarge the replica symmetric region beyond $\bar t(\beta,h)$, which is clearly impossible due to the very definition of $\bar t(\beta,h)$.

There are two important properties of the replica symmetric region that we immediately point out.

Introduce, for generic replicas $s$, $s^{\prime}$, the rescaled overlaps defined as the random variables
\begin{equation}\label{xi}
\xi_{ss^{\prime}}\equiv \sqrt{N} (q_{ss^{\prime}}-\bar q(\beta,h)),
\end{equation}
distributed according to the $\med{\dots}_{\lambda,t}$ state.
Then we have
\begin{theorem}\label{gauss}
For any $(\lambda,t)$ in the replica symmetric region, in the infinite volume limit, the $\xi$ become centered Gaussian random variables, whose covariances can be explicitely calculated. These covariances may become singular only along the boundary line $\bar\lambda$, or at the boundary point $\bar t(\beta,h)$, according to the various situations analyzed in the following.
\end{theorem}
The proof is very simple. In fact, it is a simple adaptation of the same procedure exploited in \cite{GTcentral}, and \cite{T}, for the case where $\lambda=0$. The effective calculation of the covariances exploits methods introduced in \cite{Gsum}, \cite{GTcentral}, and \cite{T}. We sketch the extension to the case where $\lambda>0$ in the next section 5.

In the replica symmetric region, we have not only the convergence of the $\tilde\alpha_{N}(\lambda,t)$ auxiliary function to $\alpha_{SK}(\beta,h)$, but also a stronger statement.
\begin{theorem}\label{fi}
For any $(\lambda,t)$ in the replica symmetric region, with the possible exclusion of the end point $\bar t(\beta,h)$, the infinite volume limit
\begin{equation}\label{fi'}
\lim_{N\to\infty} N(\tilde\alpha_{N}(\lambda,t)-\alpha_{SK}(\beta,h))\equiv \phi(\lambda,t)
\end{equation}
is well defined and can be explicitly given in terms of integrals of fluctuations as follows
\begin{equation}\label{fi''}
\phi(\lambda,t)=-\frac{\beta^{2}}{4}\int_{0}^{t} A(t^{\prime}) d t^{\prime}+\frac{\beta^{2}}{4}\int_{0}^{\lambda}A(\lambda^{\prime},t) d \lambda^{\prime},
\end{equation}
where
\begin{equation}\label{A}
A(t^{\prime})\equiv \lim_{N\to\infty} \med{\xi_{12}^{2}}_{t^{\prime}},\ \ \ A(\lambda^{\prime},t)\equiv \lim_{N\to\infty}\med{\xi_{12}^{2}}_{\lambda^{\prime},t}.
\end{equation} 
The function $\phi$ could take the value $\infty$ at the boundary line $\bar\lambda$, 
according to the various situations analyzed in the following.
\end{theorem}

The proof is very simple. In fact, we recall that $\tilde\alpha_{N}(0,t)\equiv\alpha_{N}(t)$, and write
\begin{equation}\label{fisum}
N(\tilde\alpha_{N}(\lambda,t)-\alpha_{SK}(\beta,h))\equiv
-N(\alpha_{SK}(\beta,h)-\alpha_{N}(t))+N(\tilde\alpha_{N}(\lambda,t)-\tilde\alpha_{N}(0,t)). 
\end{equation} 
Then we exploit (\ref{dalfaSK}) and (\ref{dlambdaalfatilde'}). By integration we get
\begin{eqnarray}\label{int} 
N(\alpha_{SK}(\beta,h)-\alpha_{N}(t))&=&
\frac{\beta^{2}}{4}\int_{0}^{t}\med{\xi_{12}^{2}}_{t^{\prime}} d t^{\prime}, \\
\label{int'}
N(\tilde\alpha_{N}(\lambda,t)-\tilde\alpha_{N}(0,t))&=&
\frac{\beta^{2}}{4}\int_{0}^{\lambda}\med{\xi_{12}^{2}}_{\lambda^{\prime},t} d \lambda^{\prime}.
\end{eqnarray}
By taking the infinite volume limit we get (\ref{fi''}), and Theorem \ref{fi} is proven.

It is also convenient to introduce the level lines of the function $\phi$. In fact, we follow the procedure exploited in section 2, by starting with the level lines for the the difference $\tilde\alpha_{N}(\lambda,t)-\alpha_{SK}(\beta,h))$. It is immediately recognized, as in section 2, that, for finite $N$, the corresponding velocity field is given by
\begin{equation}\label{vi'}
v(\lambda,t)=-1+2\frac{\med{(q_{13}-\bar q)^{2}}_{\lambda,t}}{\med{(q_{12}-\bar q)^{2}}_{\lambda,t}}\equiv-1+2\frac{\med{\xi_{13}^{2}}_{\lambda,t}}{\med{\xi_{12}^{2}}_{\lambda,t}}.
\end{equation}  
In the replica symmetric region, fluctuations can be completely controlled. Therefore, in the infinite volume limit, we have that the velocity field of the level lines for the limiting $\phi$ function are given explicitly by
\begin{equation}\label{vilimit}
v(\lambda,t)=-1+2\frac{D(\lambda,t)}{A(\lambda,t)},
\end{equation}
where $A(\lambda,t)$ is defined in (\ref{A}), and
\begin{equation}\label{D}    
D(\lambda,t)=\lim_{N\to\infty}\med{\xi_{13}^{2}}_{\lambda,t}.
\end{equation}
Notice that $A(\lambda,t)$ and $D(\lambda,t)$ are different in general, because the $\lambda$ coupling destroys the $(2-3)$ symmetry. They are explicitly calculated by following the method outlined in section 5. Of course, at $\lambda=0$, we have $A(0,t)\equiv D(0,t)\equiv A(t)$. In particular $v(0,t)\equiv1$, for any $t<\bar t(\beta,h)$, where $A(t)$ is surely not singular.   

Let us give a simple example. We consider the annealed case, where $h=0$, $\bar q=0$. Here everything can be explicitly calculated. In particular, the methods of \cite{GTquadratic} give $\bar t(\beta)=1$ if $\beta\le1$, and $\bar t(\beta)=\beta^{-2}$
if $\beta>1$. The annealed region is given by $0\le t\le \bar t(\beta)$, $0\le\lambda\le \bar\lambda(t)$, where $\bar\lambda(t)=\beta^{-2} - t$. In the annealed region, by following the methods outlined in section 5, we find that the variances are given by
\begin{equation}\label{AD}
A(t)=1/(1-t\beta^{2}),\ \ \  D(\lambda,t)\equiv A(t),\ \ \ A(\lambda,t)=1/(1-(\lambda+t)\beta^{2}). 
\end{equation}
Notice that $A(\lambda,t)$ becomes infinite on the $\bar\lambda$ line, where $v(\bar\lambda(t),t)=-1$.
In general we have for the velocity field, from (\ref{vilimit}) and (\ref{AD}),
\begin{equation}\label{viannealed}
v(\lambda,t)=-1+2\frac{D(\lambda,t)}{A(\lambda,t)}\equiv 1+2\frac{\lambda\beta^{2}}{1-t\beta^{2}}
\end{equation}
The level lines are immediately found by integration of the streaming equation $\dot \lambda(t)=v(\lambda(t),t)$. Let $\lambda_{0}$ be a parameter with the condition $\lambda_{0}\le\bar\lambda_{0}\equiv\beta^{-2}$. Then the family of level lines is given by
\begin{equation}\label{levellines}
\lambda(t)=(\beta^{-2}-t)-(\beta^{-2}-\lambda_{0})(1-\beta^{2}t)^{2},
\end{equation}
where in any case it must be $0\le t\le1$ and $\lambda(t)\ge0$.
If $\lambda_{0}\ge0$, then the trajectories start from $\lambda_{0}$ at $t=0$. If $\lambda_{0}<0$ then the trajectories start from $\lambda=0$ at $t=\lambda_{0}/(1-\beta^{2}\lambda_{0})$.  
The value of the function $\phi$, as defined in (\ref{fi'}), can be easily found from (\ref{fi''}) in the form
\begin{equation}\label{fiannealed}
\phi(\lambda,t)=\frac{1}{4}\log\frac{1}{1-\lambda_{0}\beta^{2}},
\end{equation}
where $\lambda_{0}$ is the parameter characterizing the trajectory passing through $(\lambda,t)$.

In this simple completely solvable annealed case, we can point out the key phenomenon appearing on the $(\lambda,t)$ plane. 
Assume $\beta>1$. Then we know $\bar t(\beta)\equiv\beta^{-2}<1$.
The annealed region is the triangle $0\le t\le\bar t(\beta)$, $0\le\lambda\le\beta^{-2}-t$. At the end point $t=\bar t(\beta)$, $\lambda=0$, the variance $A(t)$ becomes infinite, the function $\phi(0,\bar t(\beta))$ is not well defined, and moreover all trajectories have there an osculating focus with common slope $v(0,\bar t(\beta))=-1$.

We will show in the following that this kind of picture holds qualitatively also in the general replica symmetric case.    

\section{Calculation of overlap fluctuations}

Overlap fluctuations have been calculated in the infinite volume limit in \cite{Gsum} and \cite{GTcentral}, see also \cite{T}.

Firstly, we consider the case $\lambda=0$. The initial conditions at $t=0$ follow easily from the central limit theorem. In fact we have
\begin{eqnarray}\label{AB0}
A_{0}&\equiv&\lim_{N\to\infty}\med{\xi_{12}^{2}}_{0}\equiv1-\bar q^{2},\ \ \
B_{0}\equiv\lim_{N\to\infty}\med{\xi_{12}\xi_{13}}_{0}\equiv\bar q-\bar q^{2},\\
\label{C0}
C_{0}&\equiv&\lim_{N\to\infty}\med{\xi_{12}\xi_{34}}_{0}\equiv\int\tanh^{4}(\beta(h+\sqrt{\bar q(\beta,h)}z)) d\mu(z)-\bar q^{2}. 
\end{eqnarray}
It is also convenient to define
\begin{equation}\label{yz0}
y_{0}\equiv A_{0}-2B_{0}+C_{0},\ \ \ z_{0}\equiv A_{0}-4B_{0}+3C_{0}. 
\end{equation}
It is immediately recognized that $y_{0}$ has the value already given in (\ref{y0}). Moreover, the bound $z_{0}\le y_{0}$ holds.

In the replica symmetric interval, in the infinite volume limit, the $\xi$ become centered Gaussian random variables, with explicitly given covariances (see \cite{Gsum}, \cite{GTcentral}, and \cite{T}) according to
\begin{eqnarray}\label{At}
A(t)\equiv&\lim_{N\to\infty}\med{\xi_{12}^{2}}_{t}&=3y(t)-2z(t)+w(t),\\
\label{Bt}
B(t)\equiv&\lim_{N\to\infty}\med{\xi_{12}\xi_{13}}_{t}&=\frac{3}{2}y(t)-\frac{3}{2}z(t)+w(t),\\
\label{Ct}
C(t)\equiv&\lim_{N\to\infty}\med{\xi_{12}\xi_{34}}_{t}&=y(t)-z(t)+w(t), 
\end{eqnarray}
where
\begin{equation}\label{yzw}
y(t)=\frac{y_{0}}{1-t\beta^{2}y_{0}},\ \ \ z(t)=\frac{z_{0}}{1-t\beta^{2}z_{0}},\ \ \ w(t)=\frac{3C_{0}-2B_{0}}{(1-t\beta^{2}z_{0})^{2}}. 
\end{equation}

In order to calculate the covariances for $\lambda\ge0$ we follow a very simple strategy. For any $t$ in the replica symmetric interval, we take $A(t),B(t),C(t)$ as initial conditions, and consider the streaming equations in $\lambda$. These are easily solved in the infinite volume limit. There is a mild complication. Since the symmetry between replicas is explicitly broken by the $\lambda$ couplings, there are more basic covariances to be taken into account. Let us consider some examples.

Define, in agreement with (\ref{A}),
\begin{equation}\label{AC}
A(\lambda,t)\equiv \lim_{N\to\infty}\med{\xi_{12}^{2}}_{\lambda,t},\ \ \ C(\lambda,t)\equiv \lim_{N\to\infty}\med{\xi_{12}\xi_{34}}_{\lambda,t},
\end{equation}
with the initial conditions $A(0,t)=A(t)$, $C(0,t)=C(t)$. At fixed $t$, the streaming equations are
\begin{eqnarray}\label{dlambdaA}
\frac{d}{d\lambda}\med{\xi_{12}^{2}}_{\lambda,t}&=&\frac{\beta^{2}}{2}\med{\xi_{12}^{2}(\xi_{12}^{2}-\xi_{34}^{2})}_{\lambda,t},\\
\label{dlambdaC}
\frac{d}{d\lambda}\med{\xi_{12}\xi_{34}}_{\lambda,t}&=&\frac{\beta^{2}}{2}\med{\xi_{12}\xi_{34}(\xi_{12}^{2}+\xi_{34}^{2}-\xi_{56}^{2}-\xi_{78}^{2})}_{\lambda,t}.
\end{eqnarray}
Now we go to the infinite volume limit, take into account the Gaussian nature of the resulting fluctuations, and exploit the partial symmetry among replicas. We find the equations
\begin{equation}\label{AC'}
A^{\prime}(\lambda,t)=A(\lambda,t)^{2}-C(\lambda,t)^{2},\ \ \
C^{\prime}(\lambda,t)=2A(\lambda,t)C(\lambda,t)-2C(\lambda,t)^{2}, 
\end{equation}
where the prime denotes the partial derivative with respect to the rescaled parameter $\lambda\beta^{2}$. These equations, with the given initial conditions, are easily integrated in the form
\begin{eqnarray}\label{A''}
A(\lambda,t)&=&\frac{A(t)-C(t)}{1-\lambda\beta^{2}(A(t)-C(t))}+C(\lambda,t),\\
\label{C''}
C(\lambda,t)&=&\frac{C(t)}{(1-\lambda\beta^{2}(A(t)-C(t)))^{2}}.
\end{eqnarray}
Analogous procedure can be followed for the other covariances. For example, for
\begin{equation}\label{BDE}
B(\lambda,t)\equiv \lim_{N\to\infty}\med{\xi_{12}\xi_{13}}_{\lambda,t},\ \ \ D(\lambda,t)\equiv \lim_{N\to\infty}\med{\xi_{13}^{2}}_{\lambda,t},\ \ \ E(\lambda,t)\equiv \lim_{N\to\infty}\med{\xi_{13}\xi_{56}}_{\lambda,t}, 
\end{equation}
we find the equations
\begin{equation}\label{BDE'}
B^{\prime}=AB+BC-2CE,\ \ \ D^{\prime}=2B^{2}-2E^{2},\ \ \ E^{\prime}=2BC+AE-3CE, 
\end{equation}
which are easily solved by taking into account the
initial conditions $B(0,t)=B(t)$, $D(0,t)=A(t)$, $E(0,t)=C(t)$.

We end this section with the following qualitative estimates, which are useful for the investigation of the properties of the level lines introduced in the previous section.

\begin{theorem}\label{vibigger}
The following estimates hold
\begin{equation}\label{ADv}
A(\lambda,t)\le \frac{A(t)}{1-\lambda\beta^{2}A(t)},\ \ \
D(\lambda,t)\ge A(t),\ \ \ v(\lambda,t)\ge 1-2\lambda\beta^{2}A(t).   
\end{equation}
\end{theorem}
The proof is simple. By neglecting the $C^{2}$ term in the first equation in (\ref{AC'}), we obtain $A^{\prime}\le A^{2}$, which gives the first inequality in (\ref{ADv}) upon integration. On the other hand, by exploiting (\ref{BDE'}), we find
\begin{equation}\label{BmenoE}
(B-E)^{\prime}=(B-E)(A-C).
\end{equation}  
In general $A\ge C$ by Schwarz inequality, and $B\ge E$ at $\lambda=0$, because here it reduces to the obvious $B(t)\ge C(t)$. Therefore $B\ge E$ for any $\lambda$. But then, from (\ref{BDE'}), we derive $D^{\prime}\ge0$, and establish $D(\lambda,t)\ge A(t)$. Finally, the lower bound for $v$ comes from the previous bounds, and the definition (\ref{vilimit}).    
 
\section{The replica symmetric region extends up to the Almeida-Thouless line}

Now we are ready to prove Theorem \ref{maint}. Assume that 
$\beta^{2}y_{0}(\beta,h)>1$. We show that $\bar t(\beta,h)$ must have exactly the value $\bar t(\beta,h)=\beta^{-2}y_{0}(\beta,h)^{-1}$. By Toninelli argument in \cite{TAT}, we already know $\bar t(\beta,h)\le\beta^{-2}y_{0}(\beta,h)^{-1}$. Suppose $\bar t(\beta,h)<\beta^{-2}y_{0}(\beta,h)^{-1}$. Then the variance $A(t)$ would stay finite up to the point $t=\bar t(\beta,h)$ included, as the explicit solution in (\ref{At}) shows (recall that $z_{0}\le y_{0}$). As a consequence, the function $\phi$, defined in (\ref{fi'}), would be well defined at the point $(\lambda=0, t=\bar t(\beta,h))$, due to the integral representation (\ref{fi''}) taken at $\lambda=0$. Here we find an immediate contradiction. In fact, due to the condition (\ref{focus'}), in the point $(\lambda=0, t=\bar t(\beta,h))$ all level lines should converge. But each of them carries its own value for the function $\phi$, depending on the initial condition $\lambda_{0}$. For example, along the trajectory starting from $\lambda=0$ at $t=0$, we must have $\phi=0$. As a general fact, the level lines can not focus on a point where $\phi$ is well defined. We conclude that $\bar t(\beta,h)<\beta^{-2}y_{0}(\beta,h)^{-1}$ is impossible. We must have $\bar t(\beta,h)=\beta^{-2}y_{0}(\beta,h)^{-1}$. Therefore, Theorem \ref{maint} is proven, together will all results of this paper.

We end this section with an argument showing the unwillingness of the level lines to focus in a point where $A(t)$ is not singular. Let us recall that we have proven the bound $v(\lambda,t)\ge1-2\lambda \beta^{2} A(t)$ (see (\ref{ADv})), for the velocity field of the level lines. Now assume that $A(t)$ is finite along the whole replica symmetric interval, so that $A(t)\le a$, for some positive constant $a$. Then we would have $v(\lambda,t)\ge1-2\lambda \beta^{2} a$. This can be easily integrated. For example, we would find that the level line starting from $\lambda=0$ at $t=0$, should stay above the line 
\begin{equation}\label{sopra}
\bar\lambda(t)=(1-\exp(-2t\beta^{2}a))/2\beta^{2}a,
\end{equation}
which is clearly incompatible with (\ref{focus'}).

In conclusion, our general argument shows that, on the $(\beta,h)$ plane, the boundary between the replica symmetric region and the broken replica symmetry region is characterized by a singularity in the rescaled overlap fluctuations, exactly as happens in the well known case of vanishing magnetic field. Therefore the transition is exactly at the Almeida-Thouless line.

It is interesting to compare our results with the characterization of the transition line given by the very clever method of Talagrand \cite{T}. Talagrand exploits broken replica symmetry bounds for a system made by two coupled replicas. As a result, the transition line is given in terms of the onset of the instability of the replica symmetric solution in the frame of a one step breaking of the replica symmetry in the Parisi representation. Therefore, the problem of comparing the Talagrand transition line with the Almeida-Thouless line is open. In our method, we do not employ any information coming from the breaking of the replica symmetry, but exploit only the peculiar properties of the quadratic replica coupling. It is also important to remark that Talagrand method has been extended to give the proof of the validity of the Parisi representation for the free energy in the infinite volume limit \cite{Topus}. In next section we deal with the problem of extending our method to more general situations.

\section{Conclusion and outlook for future developments}

We have seen that the control of the rescaled overlap fluctuations and the peculiar properties of the quadratic replica coupling allow to show that replica symmetry holds up to the Almeida-Thouless line. No information coming from the pattern of the spontaneous replica symmetry breaking phenomenon is necessary. 

In its very essence, this method allows to isolate an important parameter ruling the phenomenon. The replica symmetric representation holds, provided the parameter $\beta^{2}(A_{0}-2B_{0}+C_{0})$ is less or equal to 1. Therefore, the ruling parameter is given in terms of fluctuations at zero coupling!

It is important to remark that the flow equations for fluctuations do not depend on the particularly chosen representation. The solutions do depend on the representation, only through the initial conditions $(A_{0}, B_{0}, C_{0})$.

Therefore, it is possible to extend our method also to other representations. In particular, in the frame of the Aizenman-Sims-Starr extended variational principle \cite{ASS}, we can take the representation based on Derrida-Ruelle-Parisi probability cascades (see for example \cite{GLes}). Here sum rules also hold, and we can consider a generalized quadratic replica coupling between replicas. We plan to report on these problems in a future publication.

 \vspace{.5cm}
{\bf Acknowledgments}

We gratefully acknowledge useful conversations with Dmitry Panchenko,  Michel Talagrand, and Fabio Toninelli. Their constructive criticism was very useful in the improvement of the presentation of the results contained in this paper. The strategy exploited in this paper grew out from a 
systematic exploration of comparison and interpolation methods, developed in 
collaboration with Fabio Toninelli, and Luca De Sanctis.

This work was supported in part by MIUR 
(Italian Minister of Instruction, University and Research), 
and by INFN (Italian National Institute for Nuclear Physics).

\end{document}